\documentclass[]{raa}
\usepackage{deluxetable}
\usepackage{amssymb}            % referee version: for submission
\usepackage{graphicx,times}
\usepackage{natbib}
\usepackage{color}

\providecommand{\bm}[1]{\mbox{\bol
dmath$#1$\unboldmath}}

\begin{document}
%\SetRunningHead{H.-L. Li
%ET AL.}{THE SOLAR-TYPE BINARY OO AQUILAE}
%\Received{~~~~/~~/~~}%{yyyy/mm/dd}
%\Accepted{~~~~/~~/~~}%{yyyy/mm/dd}

\title{Low-mass and High-mass Supermassive Blackholes In Radio-Loud AGNs Are Spun-up in Different Evolution Paths}

% \volnopage{ {\bf 2009} Vol.\ {\bf 9} No. {\bf 10}, 1103--1118}
\volnopage{}
	\setcounter{page}{1}

	\author{J. Wang	
	\inst{1,2,3}
	\and M. Z. Kong\inst{4}
       \and S. F. Liu\inst{2}      
       \and D. W. Xu\inst{2,3}
	\and Q. Zhang\inst{4}
        \and J. Y. Wei\inst{2,3}
	}
 \institute{Guangxi Key Laboratory for Relativistic Astrophysics, School of Physical Science and Technology, Guangxi University, Nanning 530004, China; {wj@nao.cas.cn} 
  \and Key Laboratory of Space Astronomy and Technology, National Astronomical Observatories,
Chinese Academy of Sciences, Beijing 100012, China
\and School of Astronomy and Space Science, University of Chinese Academy of Sciences, Beijing, China
\and Department of Physics, Hebei Normal University, No. 20 East of South 2nd Ring Road, Shijiazhuang, 050024, China; {confucious\_76@163.com}\\
%	University of Chinese Academy of Sciences,
%Beijing 100049, China; \\
%   \and
%	School of Physics and Electronic Information, Huaibei
%Normal University,235000 Huaibei, Anhui Province, China;\\
%   \and
%	School of Physics and Electronic Information, Huaibei
%Normal University,235000 Huaibei, Anhui Province, China;
%
%\vs \no
%	{\small Received * ; accepted *}
}

\abstract{
How Supermassive Blackholes (SMBHs) are spun-up is a key issue of modern astrophysics.
As an extension of the study in Wang et al. (2016), we here address the issue by comparing the host galaxy properties of nearby ($z<0.05$) radio-selected Seyfert 2 galaxies. 
With the two-dimensional bulge+disk decompositions for the SDSS $r$-band images, 
we identify a dichotomy on various host galaxy properties for the radio-powerful SMBHs. 
By assuming the radio emission from the jet reflects a high SMBH spin, which stems from the well-known BZ mechanism of jet production, 
high-mass SMBHs (i.e., $M_{\mathrm{BH}}>10^{7.9}M_\odot$) have a preference for being spun-up in classical bulges,  and low-mass SMBHs (i.e., $M_{\mathrm{BH}}=10^{6-7}M_\odot$) in pseudo-bulges. This dichotomy suggests and confirms that high-mass and low-mass SMBHs are spun-up in different ways, i.e., a major ``dry'' merger and a secular evolution. 
}
%\KeyWords{ binaries: close --- binaries: eclipsing --- stars: individual (OO Aql)} }

 \authorrunning{Wang et al. }            %author_head in even pages
 \titlerunning{Spinning-up of Supermassive Blackholes}  % title_head in odd pages
 \maketitle

\section{Introduction}

Both merger and secular evolutionary scenarios have been proposed to understand the growth of
supermassive blackholes (SMBHs) located at the \bf centers of host galaxies, which stem \rm from the
widely accepted conception of co-evolution
of active galactic nuclei (AGNs) and their host galaxies (e.g.,
Heckman \& Best 2014; Alexander \& Hickox 2012; Sanders et al. 1988).
Although a high fraction of merger is found in luminous quasars and ultra-luminous infrared galaxies (ULIRGs)
(e.g., Liu et al. 2008; Mainieri et al. 2011; Treister et al. 2012; Veilleux et al. 2009; Hao et al. 2005),
the studies from Sloan Digital Sky Survey (SDSS) clearly suggest that the growth of local less-massive SMBHs
are mainly resulted from gas accretion occurring in small host galaxies (see a review in Heckman \& Best 2014),
which implies a prevalence of a disk-like bugle (i.e., a pseudo-bugle, e.g, Kormendy \& Kennicutt 2004)
for these host galaxies.
In fact, observations with high spatial resolution reveal a pseudo-bugle in the galaxies of some narrow-line Seyfert 1
galaxies (NLS1s) that are believed to have less-massive SMBHs and high Eddington ratios (e.g., Zhou et al. 2006; Ryan et al. 2007;
Mathur 2000; Orban de Xivry et al. 2011; Mathur et al. 2012). Wang et al. (2016)
recently claimed that for less-massive SMBHs powerful radio emission is favored for occurring in pseudo-bugles, which
implies that the less-massive SMBHs are spun up
by a gas accretion due to the significant
disk-like rotational dynamics of the host galaxy  in the secular evolution scenario.

Are high-mass SMBHs spun up in the same way, or not? Although there was accumulating observational evidence supporting that
radio-loud quasars could be spun up by a BH-BH merger (e.g., Laor 2000; Best et al. 2005;
Chiaberge \& Marconi 2011; Chiaberge et al. 2015),
a comparison study between low-mass and high-mass SMBHs is still rare.  
The morphology of the host galaxies of quasars is in fact hard to be
studied because the host galaxies are overwhelmed by the luminous emission from the central SMBH accretion,
even though previous studies indicate that
the bulge morphology keep the evolutionary information well (e.g., Kormendy \& Kennicutt 2004; Kormendy \& Ho 2013).   
Barisic et al. (2019) recently reveals  a higher radio-loud fraction in elliptical 
galaxies with larger mass and higher stellar velocity dispersion than in disk galaxies with 
smaller mass and lower velocity dispersion, which implies that the star formation in the 
elliptical galaxies is suppressed by the feedback energy deposited by AGN's jet.

In this paper, by following our previous study in Wang et al. (2016), we attempt to explore the role of
both merger and secular evolution scenarios  on the spinning-up of a SMBH
in a sample of nearby radio-selected Seyfert 2 galaxies with powerful radio emission, in which 
the bulge morphology of high-mass SMBHs are compared to that of  low-mass SMBHs.
Our study is based on the well-known Blandford-Znajek model (Blandford \& Znajek 1977) 
in which the observed powerful jet is resulted from an extraction of the rotational energy of 
the central SMBH. In fact, Martinez-Sansigre \& Rawlings (2011) indicates that the efficiency with which 
the jet is produced is required to increase with SMBH's spin to reproduced the observe quasar's 
``radio-loudness'' range, although a direct correlation between radio power and the measured spin has not 
been found in AGNs (see Reynolds 2019 for a recent review). The lack of the correlation implies an 
importance of the strength and geometry of the magnetic field in the production of a jet.

The paper is organized as follows. Section 2 presents the sample selection and analysis.
The results and discussions are given in Section 3.
A $\Lambda$CDM cosmology with parameters $H_0=70\ \mathrm{km\ s^{-1}\ Mpc^{-1}}$,
$\Omega_{\mathrm{m}}=0.3$, and $\Omega_\Lambda=0.7$ is adopted throughout the paper.

\section{Sample and Analysis}

\subsection{Sample selection}

A sample of nearby radio-selected Seyfert 2 galaxies is used in the current study, which
is consist of two sub-samples with small and large $M_{\mathrm{BH}}$.
The sub-sample with small $M_{\mathrm{BH}}$ comes from our previous study, in which Wang et al. (2016) selected a
sample of radio-selected nearby ($z<0.05$) ``pure'' Seyfert 2 galaxies with small $M_{\mathrm{BH}}$ ($10^{6-7}\mathrm{M_\odot}$)
by cross-matching the value-added SDSS Data Release 7 Max-Planck Institute for Astrophysics/Johns Hopkins University
(MPA/JHU) catalog (see Heckman \& Kauffmann 2006 for a review) with the FIRST survey catalog
(Becker et al. 2003). Briefly speaking, not only the three widely used Baldwin-Phillips-Terlevich diagnostic diagrams (e.g.,
Veilleux \& Osterbrock 1987), but also the [OIII]$\lambda5007$/[OII]$\lambda3727$ line ratio corrected by the local extinction (Heckman et al. 1981) is
used to remove starforming galaxies, composite galaxies and LINERs (e.g, Kewley et al. 2001, 2006).
The $M_{\mathrm{BH}}$ of each galaxy is obtained from the measured velocity dispersion $\sigma_\star$ of the bulge
through the well-calibrated $M_{\mathrm{BH}}-\sigma_\star$ relationship (Magorrian et al. 1998; McConnell \& Ma 2013 and references therein)
$\log(M_{\mathrm{BH}}/M_\odot)=(8.32\pm0.05)+(5.64\pm0.32)\log(\sigma_\star/200\ \mathrm{km\ s^{-1}})$
that is valid for $M_\mathrm{BH}$ in a range of $10^{6-10}M_\odot$.
Although there is evidence that pseudo-bulges deviate from the $M_{\mathrm{BH}}-\sigma_\star$ relationship 
established in classical bulges (e.g., Kormendy \& Ho 2013),  we argue that the deviation is not a serious issue for 
the current study because the $M_{\mathrm{BH}}-\sigma_\star$ relationship is only used by us to select 
SMBHs at both high-mass and low-mass ends.

We selected a sub-sample with large $M_{\mathrm{BH}}$ by following the scheme adopted in Wang et al. (2016).
The $M_{\mathrm{BH}}$ that are obtained again from the $M_{\mathrm{BH}}-\sigma_\star$ relationship are
required to be larger than $10^{7.9}M_\odot$. This lower limit of $M_{\mathrm{BH}}$ is adopted by taking into account of
a balance between threshold and sample size.

After the selection on $M_{\mathrm{BH}}$,
the  sub-sample with large $M_{\mathrm{BH}}$ is further filtered out  according to their nuclear accretion properties.
By using the [OIII]$\lambda5007$ line luminosity as a proxy of the
bolometric luminosity (e.g., Kauffmann et al. 2003), we finally focus on the objects located within a bin of
$\log L_{\mathrm{[OIII]}}=40.5-41.5$, where the intrinsic extinction due to the host galaxy has been corrected by the standard
method based on both the Balm er decrement in the standard case B recombination and the Galactic extinction curve with $R_V=3.1$.
The used bin size is determined by a balance between
the distribution of $\log L_{\mathrm{[OIII]}}$ and the size of our finally used sample.

Finally, there are 31 objects in the sub-sample with a small $M_{\mathrm{BH}}$, and 26 in the sub-sample with a large $M_{\mathrm{BH}}$.

\subsection{Two-dimensional bulge+disk decomposition}

As the same as in Wang et al. (2016),  we model the surface brightness profile of each galaxy 
by a  linear combination of an exponential radial profile
for the disk component and a Sec profile with an index of $n_{\mathrm{B}}$
for the bulge component.  The two-dimensional bulge+disk decomposition
is performed for the $r$-band images of each of the objects listed in the large $M_{\mathrm{BH}}$ sub-sample by using the publicly available
SEX TRACTOR and KIM2D packages (Beaten \& Aunts
1996; Smart et al. 2002), except for four cases.
The decomposition is ignored for SDSS\,J081937.87+210651.4 and SDSS\,J111349.74+093510.7 because of their heavy obstruction.
The other two objects (i.e., SDSS\,J080446.40+104635.8  and
SDSS\,J130125.26+291849.4) are ignored in our decomposition since their host galaxies are
strongly disturbed due to an on-going merger of two galaxies. The SDSS $r$-band images are displayed in Figure 1 for the two objects with an on-going merger.
The seeing effect has been taken into account of by convolving the model with a
simple point-spread function described by a Gaussian profile that is determined from the field stars.
The resulted reduced $\chi^2$ is very close to unit for all the remaining 22 host galaxies.

 \begin{figure}
   \centering
   \includegraphics[width=8cm]{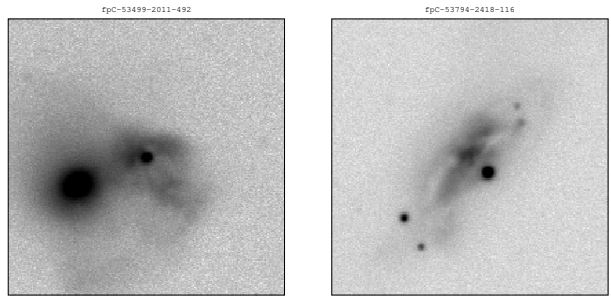}
   \caption{The two SDSS $r$-band images for SDSS\,J130125.26+291849 (the left panel) and
SDSS\,J080446.40+104635.8 (the right panel) both with a strongly disturbed profile. }
              \label{Fig1}%
\end{figure}

\section{Results and Implications}

\subsection{A Dichotomy of Pseudo-bulges and Classical Bulges for Radio-powerful Low-mass and High-mass SMBHs}

With the bulge+disk decomposition of the surface brightness,
Figure 2 reproduces the Figure 2 in Wang et al. (2016) by complementing the objects listed in the large $M_{\mathrm{BH}}$ sub-sample,
in which the modeled Sersic index $n_{\mathrm{B}}$ of the surface brightness profile of the bulge of the host galaxies
is plotted against the radio loudness $R'$ (the left panel), the rest-frame [OIII] line luminosity $L_{\mathrm{[OIII]}}$ (the middle panel),
and the rest-frame radio power $P_{\mathrm{1.4GHz}}$ at 1.4GHz (the right panel).
A $k$-correction is performed in the calculation of $P_{\mathrm{1.4GHz}}$
by adopting a universal
spectral slope $\alpha=-0.8$ ($f_\nu\propto\nu^\alpha$, Ker et al. 2012):
$P_{\mathrm{1.4GHz}}=4\pi d_L^2f_\nu(1+z)^{-1-\alpha}$,
where $d_L$ is the luminosity distance, $z$ the redshift and $f_\nu$ the observed integrated flux
density.

By combining the two traditionally used bolometric corrections: $L_{\mathrm{bol}}\approx3500L_{\mathrm{[OIII]}}$ and
$L_{\mathrm{bol}} = 9\lambda L_{\lambda}(\mathrm{5100\AA})$ (Kaspi et al. 2000; Heckman \& Best 2014),
the radio loudness $R'$ based on the [OIII] line luminosity is defined as
\begin{equation}
  \log R'=\log\bigg(\frac{P_{\mathrm{1.4GHz}}/\mathrm{W\ Hz^{-1}}}{L_{\mathrm{[OIII]}}/\mathrm{erg\ s^{-1}}}\bigg)+19.18
\end{equation}

\begin{figure}
   \centering
   \includegraphics[width=8cm]{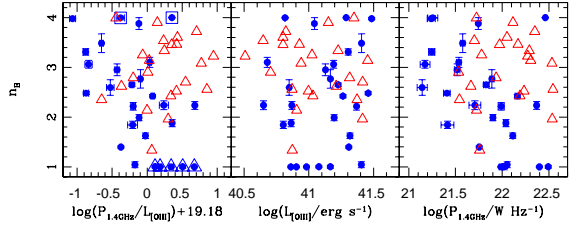}
   \caption{The modeled Sersic index $n_{\mathrm{B}}$ plotted against radio loudness $R'$ defined in Equation (1) (the left panel),
rest-frame [OIII]$\lambda$5007 line luminosity (the middle panel), and rest-frame radio power at 1.4GHz (the right panel).
The objects listed in the low-mass sub-sample are shown by the solid blue squares, and the objects listed in the
high-mass sub-sample by the open triangles. The objects with a fixed value of $n_{\mathrm{B}}$
are marked by triangles for $n_{\mathrm{B}}=1$ and by squares for $n_{\mathrm{B}}=4$.}
              \label{Fig2}%
\end{figure}

A comparison between small-$M_{\mathrm{BH}}$ and large-$M_{\mathrm{BH}}$ sub-samples in the occupation in the diagrams 
indicates that 1) almost all the objects with large $M_{\mathrm{BH}}$ are associated with a classical bulge with $n_{\mathrm{B}}>2.0$
(e.g., Kormendy \& Kennicutt 2004; Fisher \& Drory 2008); 2)
the radio-powerful (i.e., $\log R'>0$ or $\log P_{\mathrm{1.4GHz}}>21.5$) low-mass SMBHs tend to be associated with a
pseudo-bulge with $n_{\mathrm{B}}<2.0$, even though the approximation to identify pseudo-bulges by the threshold of
$n_{\mathrm{B}}$ was proposed by Gadotti (2009). To reveal the dichotomy on pseudo-bulges and classical bulges,
we perform a two-sample Gehan's generalized Wilcoxon test  on the
distributions of $n_{\mathrm{B}}$ for the radio-powerful objects with either $\log R'>0$ or  $\log P_{\mathrm{1.4GHz}}>21.5$. The statistical results are tabulated in Table 1.
Columns (2) and (3) list the probability that the two samples are drawn from the same parent population and the
corresponding $Z$-value, respectively. The average value and the corresponding standard deviation
are listed in the first row in Column (4) for the small-$M_{\mathrm{BH}}$  sub-sample, and in the 
second row for the large-$M_{\mathrm{BH}}$ sub-sample.

\begin{table}
\caption{Statistical results of two-sample Gehan's generalized Wilcoxon tests for radio-powerful SMBHs ($\log R'>0$ or $\log P_{\mathrm{1.4GHz}}>21.5$).}             % title of Table
\label{table:1}      % is used to refer this table in the text
\centering                          % used for centering table
\begin{tabular}{cccc}        % centered columns (4 columns)
\hline\hline
Parameter & $P$ &  $Z$-value & Mean \\
(1) & (2) & (3) & (4)\\
\hline
$n_{\mathrm{B}}$ & 1.2$\times10^{-3}$ & 3.242 & $1.90\pm0.29$\\
& & & $2.99\pm0.16$ \\
$\log(h_d/R_e)$ & 1.959$\times10^{-1}$ & 1.293 & $0.17\pm0.13$\\
& & & $0.42\pm0.09$ \\
$D_n(4000)$ & $5\times10^{-4}$ & 3.474 & $1.39\pm0.03$\\
& & & $1.59\pm0.04$ \\
\hline                                   %inserts single line
\end{tabular}
\end{table}

In addition to the revealed dichotomy on the Sersic index $n_{\mathrm{B}}$, the 
discrepancy between the  small-$M_{\mathrm{BH}}$  and  large-$M_{\mathrm{BH}}$ subsamples
can be further verified in Figure 3  for the radio-powerful SMBHs. 
The figure plots radio loudness $R'$ as a function of
the stellar population age (the upper panel) as assessed by the lick 4000\AA\ break index defined as
$D_n(4000)=\int_{4000}^{4100}f_\lambda d\lambda/\int_{3850}^{3950}f_\lambda d\lambda$
(e.g., Bruzual \& Charlot 2003; Coelho det al. 2007 and references therein)
and the scalelength ratio $h_d/R_e$ between discs and bulges (the lower panel).
As revealed in Wang et al. (2016), the radio-powerful low-mass SMBHs (i.e., $\log R'>0$ and $n_{\mathrm{B}}<2$) are associated
with young stellar populations with $D_n(4000)<1.6$.  While, the high-mass counterparts (i.e., $\log R'>0$ and $n_{\mathrm{B}}>2$) found to be associated with both young and old stellar populations in the current study.
Compared to the radio-powerful high-mass SMBHs, the low-mass counterparts tend to have smaller $h_d/R_e$ ratio.
In fact, by performing 2-dimensional bulge+disk decomposition for a large sample of 1000 galaxies from SDSS,
Gadotti (2009) indicates that compared to the classical bulges, the pseudo-bulges tend to have younger stellar population and
higher $R_e/h_d$ ratio at the same B/T ratio, even though the author instead separates pseudo-bulges and
classical bulges in terms of the $\langle\mu_e\rangle-R_e$ relation firmly established in elliptical galaxies
(i.e., the Kormendy relation, Kormendy 1977). The statistical results based on the same two-sample test are again
listed in Table 1.

  \begin{figure}
   \centering
   \includegraphics[width=8cm]{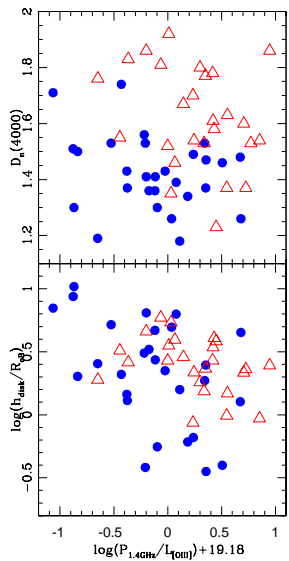}
   \caption{\it Upper panel: \rm the measured stellar population ages as assessed by the $D_n(4000)$ index is plotted as a function
of radio loudness $R'$. The symbols are the same as in Figure 2. \it Lower panel: \rm the same as the upper panel but for
the scalelength ratio between discs and bulges $h_d/R_e$.}
              \label{Fig2}%
   \end{figure}

\subsection{Merger versus Secular Evolutions: A Dichotomy of Spinning-up Mechanisms}

It is widely accepted that the SMBH's powerful radio emission is likely generated by an energy extraction from BH's spin
through the Blandford-Znajek (BZ) mechanism\footnote{An alternative is the
Blandford-Payne mechanism in which the observed powerful radio emission is resulted from
an energy extraction from a disk wind (e.g., Blandford \& Payne 1982; Wang et al. 2003; Cao 2016)}
(e.g., Blandford \& Znajek 1977, Chiaberge \& Marconi 2011; Chisellini et al. 2014).
In the model, the power of the jet $L_{\mathrm{jet}}$ is predicted to be  
$L_{\mathrm{jet}}\propto j^2B_p^2$, where $j$ is the dimensionless black hole spin and 
$B_p$ is the poloidal magnetic field strength at the horizon of the SMBH (e.g., Thorne et al. 1986; 
Meier 2001; Koide et al. 2002; Daly 2009, 2016). 
The validation of the BZ mechanism for the jet production is supported by some numerical simulations  and observations (e.g., Hawley \& Krolik 2006;
Sadowski et al. 2015; Martinez-Sansigre \& Rawlings  2011). By assuming the BZ mechanism, 
the clearly revealed dichotomy on the bugle morphology therefore predicates a profound dichotomy on the spinning-up mechanisms in
low-mass and high-mass SMBHs. We argue that the dichotomy on spinning-up mechanisms is related with the two types of
evolutionary scenario, which is described as follows.

On the one hand, a low-mass SMBHs is more likely spun-up
within a pseudo-bugle with a significant disk-like rotational dynamics.  The  pseudo-bugle can be produced in the secular evolution of a disk galaxy
possibly through either a second hump instability or a vertical
dynamical resonance
(e.g., Kormendy \& Kennicutt 2004; Fisher \& Drory 2011; Silverman et al. 2011; Kormendy \& Ho 2013; Sellwood 2014).
From the theoretical ground, a less-massive  SMBH can be spun-up efficiently by
the accreted gas through the  frame-dragging effect
that realigns the BH-disk system through the interaction between the Lense-Thirring torque and the strong
disk viscous stress (e.g., King et al. 2005, 2008; Volonteri et al. 2007; Perego et
al. 2009; Li et al. 2015), once the mass of the gas accreted onto the SMBH exceeds the
alignment mass limit $m_{\mathrm{align}}\propto a^{11/16}(L/L_{\mathrm{Edd}})^{1/8}M_{\mathrm{BH}}^{15/16}$ (King et al. 2005),
where $a=cJ/GM_{\mathrm{BH}}$ is the dimensionless angular momentum, and $L_{\mathrm{Edd}}$ the Eddington luminosity.

On the other hand, a classical bulge that is widely believed to be resulted from a major ``dry'' merger of two galaxies
(Toomre 1977) is responsible for the spinning-up of a high-mass SMBH.
A ``dry'' merger of two galaxies is argued to be the origination of a `core'' galaxy since the deficit of star light
can be resulted from an ejection of stars away from the central region during the merger (e.g., Faber et al. 1997; Kormendy et al. 2009). A spinning SMBH can be produced by the subsequent BH-BH merger if the masses of the two involved SMBH are comparable (e.g., Hughes \& Blandford 2003; Baker et al. 2006; Li et al. 2010). After the coalescences of the two SMBHs, 
a formation and a maintenance of a powerful jet results in a spinning-down due to 
an extraction of its rotational energy. 
By using the HST images with high
spatial resolution, Capetti \& Balmaverde (2006, 2007) pointed out that radio-loud AGNs tend to be associated with
a ``core'' galaxy that has a small logarithmic slope of the nuclear surface brightness profile (see also in de Ruiter et al. 2005).
In addition to the implication discussed above, the merger scenario is further supported by
the current two cases with an on-going merger (see Figure 1).
In fact, Chiaberge et al. (2015) pointed out that $\sim90\%$ radio-loud AGNs at $z>1$ are associated with an either recent or
on-going merger system.
Finally, the merger scenario is further validated by that fact that  there is accumulating evidence supporting that radio-loud AGNs
have richer environment than radio-quiet AGNs (e.g., Shen et al. 2009; Donson et al. 2010).

\section{Conclusions}

By comparing the host galaxies of a sample of nearby ($z<0.05$) radio-selected Seyfert 2 galaxies.
we identify a dichotomy on the host galaxy properties for radio-powerful SMBHs, in which
high-mass SMBHs ($>10^{7.9}M_\odot$) favor a spinning-up in classical bulges, and low-mass SMBHs ($10^{6-7}M_\odot$) in pseudo-bulges,  based on the assumption that a high spin of SMBH can be reflected 
by its powerful jet.  
We argue that high-mass and low-mass SMBHs are likely spun-up and grown up in different ways,
i.e., a major ``dry'' merger and a secular evolution, respectively.

\begin{acknowledgements}
%This work made use of data supplied by the UK Swift Science Data Centre at the University of Leicester.
%We are indebted to Dr. QIU Yu-Lei and Dr. CAO Li for
%providing original version of customed photometric reduction software.
%Also, we are grateful to WANG Yunpeng, DU Zhibo and other technical
%supports for their assistants of the observations.
%Many thanks go to Professor J.-M. Kreiner for his compiled light minimum times of this binary.
%XLP acknowledges the support of the National Natural Science Foundation
%of China (NSFC) grant 11103036 and U1331101.
   
JW \& DWX are supported by the National Natural Science Foundation of China under grants
11773036 and 11473036.
MZK is supported by NSFC Youth Foundation (No. 11303008) and by Astronomical Union Foundation under grant
No. U1831126.
This study is supported by the National Basic
Research Program of China (grant 2014CB845800), the NSFC under grants 11533003, and the Strategic
Pioneer Program on Space Science, Chinese Academy of Sciences, Grant No.XDA15052600.
The study is supported by the National Basic Research Program of China
(grant 2009CB824800).  
This study uses the SDSS archive data that was created and distributed by the Alfred P.
Sloan Foundation, the Participating Institutions, the National Science
Foundation, and the U.S. Department of Energy Office of Science.

\end{acknowledgements}

\label{lastpage}

\end{document}